\documentclass{PoS}

\newcommand{\integral}{\textit{INTEGRAL}}

\title{Narrow Line Seyfert 1s in the IBISCO sample}

\ShortTitle{IBISCO NLS1}

\author{\speaker{Manuela Molina}\\
        INAF-OAS Bologna\\
        E-mail: \email{molina@iasfbo.inaf.it}}

\author{A. Malizia\\
        INAF-OAS Bologna\\
        E-mail: \email{malizia@iasfbo.inaf.it}}
        
        \author{F. Fiore\\
        INAF-OAT Trieste\\
        E-mail: \email{fabrizio.fiore@inaf.it}}
        
        \author{C.Feruglio\\
        INAF-OAT Trieste\\
        E-mail: \email{chiara.feruglio@inaf.it}}

\abstract{We present the broad-band soft and hard X-ray spectral analysis of 8 Narrow Line Seyfert 1 galaxies extracted from the IBISCO Sample. The study also focuses 
on the properties of the NLS1 in our sample in relation to those of the IBISCO parent Seyfert population. The IBISCO sample comprises 57 AGN selected from the INTEGRAL 
IBIS AGN catalogue (in the 20-100 keV band), with z$<$0.05 and covering a wide range of luminosities, BH masses and absorption. All AGN have also measurements of the molecular gas (H$_2$) 
content of their host galaxies, through the detection of CO emission lines.
The main goals of this analysis are to accurately determine the X-ray continuum emission, investigate the presence of absorption features around 7 keV (indicative of the presence of outflows) and 
measure the bolometric luminosity in order to study the accretion parameters of the eight IBISCO NLS1, and study the accretion mechanisms and investigate the feeding and feedback cycle in these peculiar AGN. 
Preliminary results show that NLS1 tend to have higher Eddington ratios and larger molecular gas fractions than their parent Seyfert population in the IBISCO sample. 
Nuclear (AGN) vs. host galaxy properties scaling relations of NLS1 in relation to the parent Seyfert population are also explored. }

\FullConference{Revisiting narrow-line Seyfert 1 galaxies and their place in the Universe - NLS1 Padova\\
		9-13 April 2018 \\
		Padova Botanical Garden, Italy}

\begin{document}

\section{Introduction}
It is  now commonly accepted that super massive black holes (SMBH) reside at the centre of all galaxies and
that they can power the most luminous active galactic nuclei (AGN). According to observational
evidence, a tight correlation between the SMBH mass and the mass of the host galaxy bulge exists, hinting at a 
self-regulating mechanism that links the accretion-powered growth of the SMBH to star formation in the host galaxy (see e.g. \cite{marconi04}, \cite{Merloni:2008}).
This is the so-called feeding and feedback cycle of active galaxies: during the feeding phase (e.g. the luminous AGN phase), the cold and dense molecular gas responsible for star formation 
inflows towards the nucleus, causing the SMBH to grow, while during feedback,  AGN-powered winds and outflows are formed, 
causing the physics and geometry of the ISM to be changed and altering further star-formation and nuclear gas accretion.
The cycle then starts anew when cold gas begins to accrete again onto the SMBH.
The two main ingredients in this framework are therefore the molecular gas content of the host galaxies
and the AGN-driven winds; these two elements can be studied through high spatial resolution CO mapping and
deep X-ray spectroscopy. In the X-ray band, narrow emission 
line outflows with moderate (hundreds to a few thousands km/s) and high (up to 0.7c) velocities, i.e. warm absorbers and ultra-fast outflows, 
have been detected in a large fraction of AGN ( e.g.  \cite{tombesi15}), confirming that these are an efficient way 
of transporting energy from the nucleus to the host galaxy.

\section{The IBISCO Sample}
Information about cold gas in AGN is usually derived from far infrared-selected samples, which are biased towards
star-forming galaxies.  To surpass such a crucial limitation, 
we have extracted a sample of hard X-ray selected AGN from the {\integral}/IBIS AGN catalogue (\cite{Malizia_2016}). All the AGN in  this catalogue have been 
univocally associated with their X-ray counterparts and are optically identified. The sample is 
fully characterised in terms of optical identification and spectroscopy and also in terms of X-ray spectral properties.
All sources are located at  z$<$0.05, Dec$>$-20$^{\circ}$, have L$_{\rm Bol}$$>$10$^{43}$erg/s, accurate 
black hole mass estimates and cover a broad range of bolometric and Eddington luminosities,  and obscuration.
Within the IBISCO sample, a sub-sample of 8 Narrow Line Seyfert 1 Galaxies (NLS1 from now on: Mrk 110, NGC 4051, Mrk 766, NGC 4748, NGC 5506, IGR J19378-0617,
Swift J2127.4+5654 and Kaz 320) is present. NLS1 are interesting targets as they are characterised by unique properties when compared to their broad line analogues,
both in the optical (see e.g. \cite{Osterbrock_1985}) and in the X-rays, where they show strong variability
(both in flux and spectral shape) and steeper power law spectra than their broad line analogues. 
The most widely accepted explanation for these differences is that NLS1 have smaller black hole masses than normal Seyfert 1s;
however their luminosities are comparable \cite{pounds95}, suggesting that they must be
emitting at higher fractions of their Eddington luminosity and therefore they should also have higher fractional accretion rates.
A plausible scenario suggests that black holes in NLS1 have not yet been fed enough to 
become massive and are in a rapidly growing phase \cite{Mathur_2000}; if NLS1 are indeed in an early phase of black hole evolution, 
then they are key targets for studying formation and evolution of AGN. For this reason, within the IBISCO sample, NLS1 are the ideal
sources where to investigate the AGN-host galaxy scaling relations during the phase of rapid assembly of SMBH. 
In order to obtain H$_2$ measurements  for the whole sample, we have performed a survey of CO J=2-1 and J=1-0 with the IRAM 30 telescope, 
and detected either line in 75\% of the sample.

\section{NLS 1 in the IBISCO Sample: data analysis and results}

\begin{figure}
\centering
        \begin{minipage}[c]{.40\textwidth}
\centering
          \includegraphics[width=1.05\textwidth]{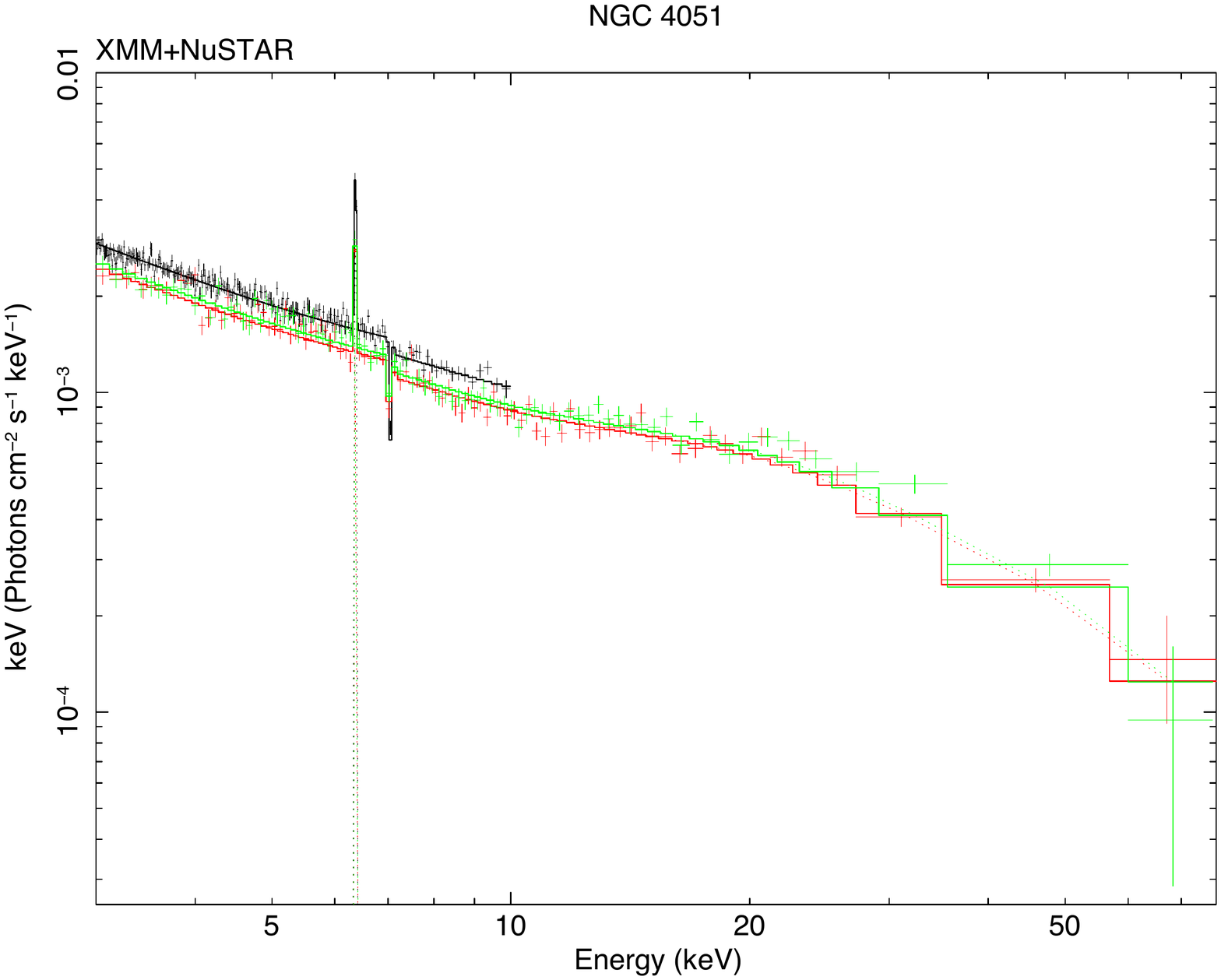}
        \end{minipage}%
        \hspace{10mm}%
        \begin{minipage}[c]{.40\textwidth}
\centering
          \includegraphics[width=1.15\textwidth]{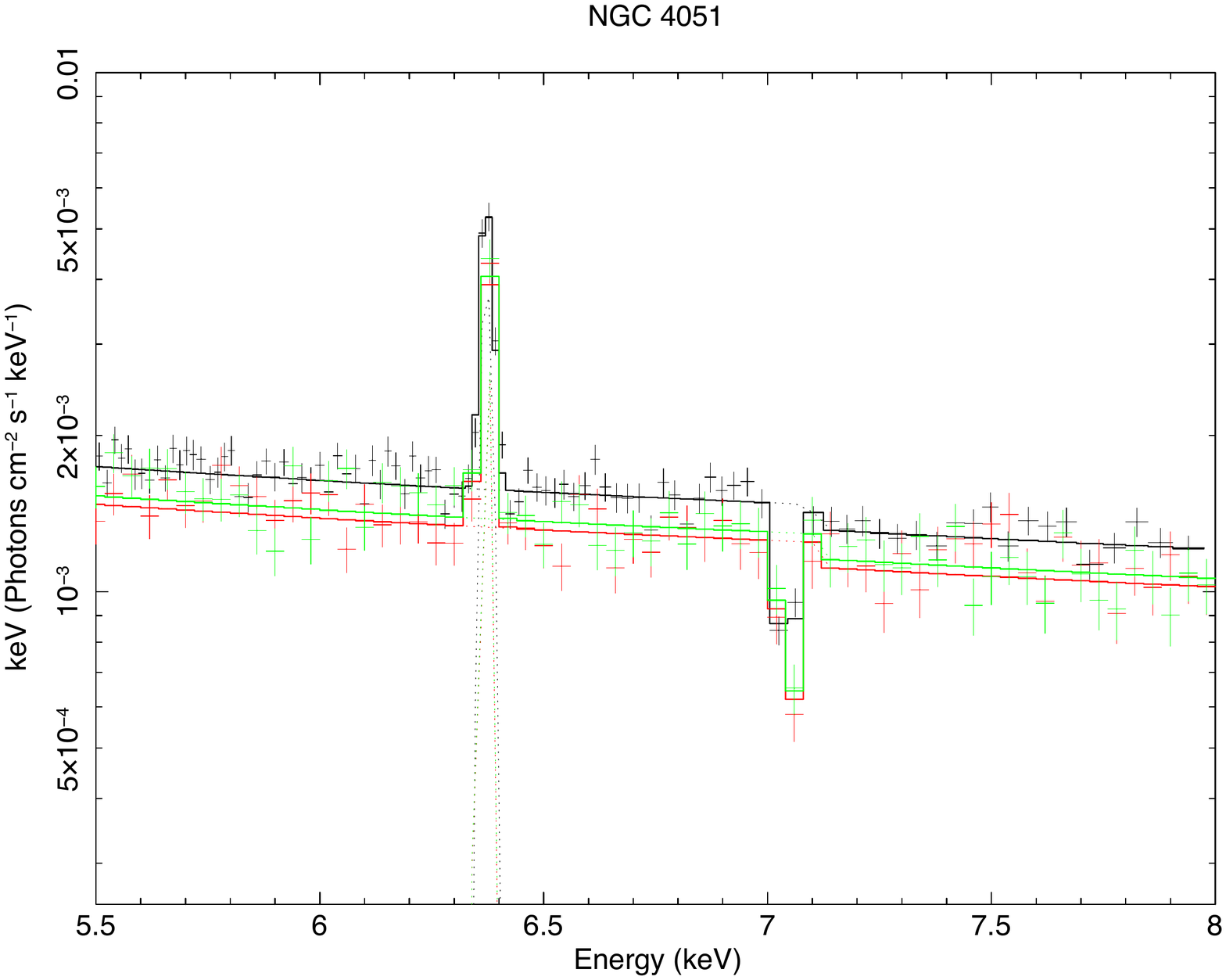}
        \end{minipage}
        \caption{\footnotesize{{\it Left panel}: 3-78 keV XMM+NuSTAR spectrum of NGC 4051. {\it Right panel}: zoom of the region around 6 keV, where the iron K$\alpha$ line is clearly 
        detected at around 6.4 keV together with an absorption feature at around 7 keV indicative of the presence of outflowing material from the innermost  region of the accretion disc.}}
        \label{ngc4051}
      \end{figure}

The eight NLS 1 in the IBISCO sample have all been analysed in the X-rays, in the range 3-78 keV (XMM plus NuSTAR data) or  3-110 keV (XMM or XRT plus INTEGRAL/IBIS and/or Swift/BAT),
preferably employing simultaneous data or observations taken in similar flux states. The broad-band spectral analysis has been conceived so that the fits are as homogeneous as possible, 
since these sources are very complex, displaying variability (either in flux, spectral shape or both) and distinctive features, above all in the soft part of their spectra. 
For this reason, we did not consider data below 3 keV, as features such as the soft excess and/or warm absorber are quite complex to  
model and their characterisation is beyond the scope of the present study. Since the main goal of this analysis is the determination of the continuum shape and X-ray luminosity of our NLSy1,
we fitted the data with a common, relatively simple phenomenological model: an exponential cut-off power law reflected from neutral material plus as many Gaussian lines 
(both in emission and in absorption) as are required by the data to model the iron line profile around 6.4 keV and any absorption feature indicative of the presence of an outflow. 
Most of the sources in our sample are well-known objects, as is the case of NGC 4051, for which we used two non-simultaneous observations (but taken in similar flux states) and fit them with our 
baseline model; the resulting fit is quite good and we were able to detect an absorption feature at around 7 keV, indicative of the presence of an outflow in the source (see figure \ref{ngc4051}).

From the broad-band spectral analysis, we derived the hard X-ray and then the bolometric luminosity for each source, in order to study their accretion parameter, expressed
in terms of Eddington ratios ($\lambda$=L$_{\rm bol}$/L$_{\rm Edd}$) and mass accretion rates  as shown in table \ref{accr}.
As expected NLS1 appear to be efficiently accreting systems, as it is evident form the derived Eddington ratios
 and mass accretion rates (both in the case of a minimally spinning black hole or in the case of a Kerr black hole). However, it must be pointed out that  
 variability in the optical band can lead to quite different measurements of black hole masses and variations in the X-ray band can also lead to
 different bolometric luminosity determinations, resulting in a large scatter in $\lambda$ values.

\begin{table*}
\centering
\scriptsize
\caption{Luminosities and accretion parameters}
\vspace{0.2cm}
\begin{tabular}{lccccccc}
\hline
{\bf Source}                &    {\bf z}    &    {\bf Log(M$_{\bf BH}$) }&  {\bf L$_{\bf bol}$}         & {\bf L$_{\bf Edd}$}    & $\bf \lambda$& {$\bf \dot{M}$ ($\bf \epsilon${\bf=0.1})}&{$\bf \dot{M}$($\bf \epsilon${\bf=0.4})}\\
                                  &                  &    (M$_{\bf \odot}$)              & (erg s$^{\bf -1}$)       & (erg s$^{\bf -1}$)   &                       &(M$_{\bf \odot}$/yr)                     & (M$_{\bf \odot}$/yr)                   \\ 
\hline
Mrk 110                     &0.035291  &                7.29              & 2.45$\times$10$^{45}$&2.46$\times$10$^{45}$&0.996             &       4.322$\times$10$^{-1}$                 & 1.080$\times$10$^{-1}$\\
NGC 4051                & 0.002336  &                6.13              & 2.51$\times$10$^{42}$&1.70$\times$10$^{44}$&0.015               &       4.422$\times$10$^{-4}$                 & 1.106$\times$10$^{-4}$\\
Mrk 766                    & 0.012929  &                6.82             &1.07$\times$10$^{44}$&8.33$\times$10$^{44}$& 0.129              &       1.886$\times$10$^{-2}$                 & 4.716$\times$10$^{-3}$\\
NGC 4748                & 0.01463    &                6.41             & 5.62$\times$10$^{43}$&3.24$\times$10$^{44}$&0.173               &       9.900$\times$10$^{-3}$                 & 2.475$\times$10$^{-3}$\\
NGC 5506                & 0.006181  &                6.94              &5.75$\times$10$^{43}$&1.10$\times$10$^{45}$& 0.052              &       1.013$\times$10$^{-1}$                 & 2.533$\times$10$^{-2}$\\
IGR J19378-0617    & 0.010587   &                6.48            &1.10$\times$10$^{44}$&3.80$\times$10$^{44}$& 0.288             &       1.930$\times$10$^{-2}$                  &  4.826$\times$10$^{-2}$\\
Swift J2127.4+5654& 0.0147        &               7.18              &2.45$\times$10$^{44}$&1.90$\times$10$^{45}$&0.129              &       4.322$\times$10$^{-2}$                 & 1.080$\times$10$^{-2}$\\
Kaz 320                   & 0.0345       &                6.36              &7.24$\times$10$^{44}$&2.88$\times$10$^{45}$&0.251              &        1.275$\times$10$^{-1}$                 & 3.189$\times$10$^{-2}$\\
\hline
\label{accr}
\end{tabular}
\end{table*}

\section*{Acknowledgements}

{\small This conference has been organized with the support of the
Department of Physics and Astronomy ``Galileo Galilei'', the 
University of Padova, the National Institute of Astrophysics 
INAF, the Padova Planetarium, and the RadioNet consortium. 
RadioNet has received funding from the European Union's
Horizon 2020 research and innovation programme under 
grant agreement No~730562. \\
M.M. acknowledges financial support from NuSTAR (NARO16)/ASI/INAF
agreement F.O. 1.05.05.95 and ASI/INTEGRAL agreement F.O. 1.05.04.20.04}.

\bibliographystyle{JHEP}
          \bibliography{mol_biblio}



\end{document}